\documentclass[a4]{iopart}
\usepackage{pst-all}
\usepackage{epsfig,graphicx,float,rotating}
\newcommand{\be}{\begin{equation}}
\newcommand{\ee}{\end{equation}}
\newcommand{\bc}{\begin{center}}
\newcommand{\ec}{\end{center}}
\newcommand{\pldv}{+\negthinspace\negthinspace\negthinspace\div}

\begin{document}

\title[Transformations that separate the center of mass motion]{All coordinates transformations that separate the center of mass kinetic energy, their group structure and geometry} 
\author{L. Fortunato} 
\address{ECT*, Strada delle Tabarelle 286, I-38050 Villazzano (TN), Italy}

\begin{abstract}
The most general coordinates transformations that allow for the exact separation of the kinetic energy operator of a quantum many-body system into total center of mass kinetic energy and internal kinetic energy are found and discussed. We find i) that the suitable transformations, depending on the number of particles, have a certain number of free parameters and this allows for the generalization of the Jacobi coordinates to a much larger class of coordinates with the same properties and ii) that there is a new, uncommon, additive group structure hidden in the transformation matrices that is connected to certain geometric properties of the set of coordinates.
\end{abstract}

The kinetic energy operator of a set of particles in quantum mechanics depends upon the Laplacian of the coordinates,
\be
T=\sum_1^n T_i =\sum_1^n -\frac{\hbar^2}{2m_i} \nabla^2_{\vec r_i}
\ee
where the summation is extended over all particles $i=\{1,\cdots,n\}$ with position vectors $\vec r_i$.
The hamiltonian of the system depends on this operator and in the study of the internal excitations of a quantum system we are faced with the fundamental problem of eliminating the kinetic energy of the center of mass from the total hamiltonian.
Several transformation of coordinates, techniques and methods exists to accomplish this task \cite{Fano,MeBr}, let me just mention the transformation to Jacobi coordinates that singles out the center of mass energy \cite{Grei}. This is a crucial step in atomic, molecular and nuclear physics. The general idea behind this is the following conversion 
\be
T=\sum_1^n T_i  \longrightarrow \sum_1^{n-1} T_i' + \frac{\vec P^2}{2M}   
\label{sep}
\ee
where $\vec P= (1/n)\sum_1^n \vec p_n$ is the linear momentum of the center of mass of the system with total mass $M=\sum_1^n m_i$.

It is generally recognized that there are several sets of Jacobi coordinates that accomplish this task \cite{Fano} and that, not only one can change the labeling of particles, but can also change the order in which particles are grouped together, thus generating a large number of possible sets of Jacobi coordinates. Then, of course, several people have thought to the problem of coordinates transformation for various reasons or applications, and they have even included masses (let me cite Ref. \cite{Smi}, where a matrix very similar to the ones I derive in the following appears), but to my knowledge the most general conditions under which one can expect center of mass separation to occur have not yet been thoroughly worked out. I must therefore apologize to all those authors whose researches are not cited in the present paper, but my intent is to look at the problem under a wider angle and specific applications, although arguably very important, are not referenced.
In addition the present analysis reveals an uncommon group structure hidden in the transformation matrices that has a connection with geometrical properties and might, very likely, open up this subject to a re-interpretation.

{\it In medias res}, let's consider coordinate transformations of the type
\be
\vec x'= A \vec x 
\ee
where the components of the vectors $\vec x, \vec x'$ are connected to each other by
\be
x'_i=\sum_{j=1}^n a_{ij} x_j \;.
\ee
It is convenient to remember that each of these components might also represent a multidimensional vector.

The relations holding among the corresponding partial derivatives are 
\be
\frac{\partial}{\partial x_k} =  \sum_{i=1}^n \frac{\partial x'_i}{\partial x_k} \frac{\partial}{\partial x'_i} =
 \sum_{i=1}^n a_{ik} \frac{\partial}{\partial x'_i}
\ee
 where we have used 
\be
\frac{\partial x'_i}{\partial x_k}=\frac{\partial}{\partial x_k}\sum_{j=1}^n a_{ij} x_j = 
\sum_{j=1}^n a_{ij} \delta_{jk}=a_{ik} \;.
\ee


We are looking for the conditions to impose on the generic coordinate transformation that bring the sum of pure second derivatives of the unprimed coordinates into the sum of pure second derivatives of the primed coordinates without introducing 
mixed derivatives. In other words for each $k=1,\cdots n$, if 
\be
\frac{\partial^2}{\partial x_k^2} =  \biggl( \sum_{i=1}^n a_{ik} \frac{\partial}{\partial x'_i} \biggr) \biggl( \sum_{j=1}^n a_{jk} \frac{\partial}{\partial x'_j} \biggr) = \sum_{i,j=1}^n a_{ik}a_{jk}  \frac{\partial^2}{\partial x'_i \partial x'_j}
\ee
is the transformation of the second derivatives, the $n(n-1)/2$ conditions to forbid the presence of mixed derivatives are 
\be
\sum_{k=1}^n a_{ik}a_{jk} =0 ~~~ \forall \{i,j \} ~~\slash i \ne j=\{1,\cdots, n\}\;.
\label{cond} 
\ee
These $n(n-1)/2$ equations form a system with $n^2$ unknowns, hence $n(n+1)/2$ matrix elements might arbitrarily be chosen.
In order to insure that the last of the coordinates always coincides with the position of the center of mass, the elements in the last row must take the form $a_{ni}=1/n, \forall i=1,\cdots,n$, and hence upon simplification the system (\ref{cond}) is replaced by
\be
\left\{
\begin{array}{cc}
\biggl\{ \sum_{k=1}^{n} a_{ik}a_{jk} =0  &  ~~\forall \{i,j \} ~~\slash i \ne j=\{1,\cdots, n-1\}\\
\biggl\{ \sum_{k=1}^{n} a_{ik}=0 &  i= \{1,\cdots, n-1\}  \\
\end{array}
\label{co1}
\right.
\ee
where the first line represents a subsystem of $(n-2)(n-1)/2$ conditions of the form (\ref{cond}) with indexes $i,j$ restricted to avoid $n$ and the second line represents $n-1$ simpler conditions.
In general the system (\ref{co1}) has $n^2-n$ unknowns, of which $(n-1)(n-2)/2$ and $(n-1)$ are set and the remaining matrix elements can be chosen arbitrarily. Therefore there are entire classes of transformations that comply with the requirement (\ref{sep}).
Oftentimes it is useful to have that at least some of the vectors to correspond with relative position vectors and this implies transformation matrices with (in some rows) only two elements different from $0$ and equal to $1$ and $-1$ respectively.

Example: for two particles, $n=2$, we have the matrix 
\be
\left(
\begin{array}{cc}
a_{11}  & a_{12}  \\
a_{21} & a_{22} 
\end{array}
\right) 
\ee
The condition (\ref{cond}) takes (twice) the form
$$ a_{11}a_{21} +a_{12}a_{22} =0 $$
that, remembering that we wish to keep $a_{21} = a_{22} =1/2 $, has only one solution $a_{12}= -a_{11}\;.$

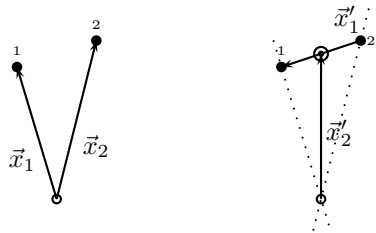
\begin{figure}[!h]
\bc
\psset{unit=1pt}
\begin{pspicture}(150,80)(0,-10)
\pscircle*[linecolor=black](0,50){2} \rput(0,56){\tiny 1}
\pscircle*[linecolor=black](30,60){2} \rput(30,66){\tiny 2}
\pscircle(15,0){2}
\psline{->}(15,0)(0,50)  \rput(2,15){$\vec x_1$}
\psline{->}(15,0)(30,60)  \rput(30,20){$\vec x_2$}

\pscircle*[linecolor=black](100,50){2} \rput(100,56){\tiny 1}
\pscircle*[linecolor=black](130,60){2}\rput(134,60){\tiny 2}
\pscircle(115,0){2}
\pscircle[doubleline=true](115,55){3}
\psline{->}(130,60)(100,50)  \rput(125,68){$\vec x_1'$}
\psline{->}(115,0)(115,55)  \rput(122,25){$\vec x_2'$}
\psline[linestyle=dotted,dotsep=3pt](118,-10)(97,60)
\psline[linestyle=dotted,dotsep=3pt](112,-12)(133,72)
\end{pspicture}
\caption{Transformation discussed in the text for 2 particles (black dots). Hollow circles represent the external reference point and double circle represents the particles' center of mass. Dotted lines are rails (see text).}
\label{ex2}
\ec
\end{figure}

Leaving aside the trivial $a_{11}= a_{12}=0$, that is unacceptable to our scopes, we have the freedom to choose the value of one of the matrix elements.  
Note that with $a_{11}= 1$ it is exactly the coordinate transformation that takes the position of two particles into the relative coordinate plus the coordinate of the center of mass, depicted in Fig. (\ref{ex2}). Any other choice would have simply altered the modulus or sign of the internal coordinate $\vec x_2'$ without changing its direction.

The most general matrix we can use for 2 particles is
\be
A_2(a)=
\left(
\begin{array}{cc}
a  & -a  \\
1/2 & 1/2 
\end{array}
\right)  
\ee
with freedom on $a$. The transformation of coordinates of Fig. (\ref{ex2}) corresponds to $A_2(1)$. Note that these matrices form an additive group under the law of composition obtained by the operation of standard matrix addition and subsequent division by two. Although there is nothing exceptionally new about this composition law (it is essentially matrix addition), I will indicate it with the $\pldv$ symbol, for the sake of brevity. Instead these matrices are {\it not} a group under usual matrix multiplication. In other words, we don't try to apply $A_2(b)$ to $A_2(a)$ and then to the column vector $(\vec x_1, \vec x_2)$: by doing so one ends up with something completely useless. But rather apply the composition $A_2(a)\pldv A_2(b)$ to the column vector: this will give again a transformation of the same type.
The meaning of this group is that all the infinite possible transformations that preserve the pure second derivatives and separate the second derivative with respect to the center of mass coordinate are essentially  equivalent, and therefore the parameter $a$ can be set to 1 without loss of information. The entire class can be 
seen geometrically as follows: in Fig. (\ref{ex2}) draw the two lines connecting the external reference point with $1$ and $2$ and call them rails; shift the vector $\vec x_1'$ along the rails, keeping it parallel to the original and adjusting the length in order to maintain its extremes on their rails. The infinite set of vectors that are obtained by this parallel expansion/contraction procedure maps one to one to the real numbers $a$.

Example: for three particles, $n=3$, the matrix is  
\be
\left(
\begin{array}{ccc}
a_{11}  & a_{12}& a_{13}   \\
a_{21} & a_{22} & a_{23} \\
a_{31} & a_{32} & a_{33} 
\end{array}
\right) 
\ee
and the conditions are
\be
\left\{
\begin{array}{cc}
a_{11} a_{21} +a_{12}a_{22}+a_{13}a_{23} & = 0   \\
a_{11} a_{31} +a_{12}a_{32}+a_{13}a_{33} & = 0   \\
a_{21} a_{31} +a_{22}a_{32}+a_{23}a_{33} & = 0 
\end{array}
\right. 
\ee
that with $a_{31} = a_{32} = a_{33}= 1/3 $
can be simplified to a system of three equations with six unknowns
\be
\left\{
\begin{array}{cc}
a_{11} a_{21} +a_{12}a_{22}+a_{13}a_{23} & = 0   \\
a_{11} +a_{12}+a_{13}  & = 0   \\
a_{21} +a_{22}+a_{23} & = 0 
\end{array}
\right. 
\label{sys3}
\ee
where two of the three equations have the form suggested in (\ref{co1}).
Therefore one has the freedom to choose three elements and deduce the others according to (\ref{sys3}).

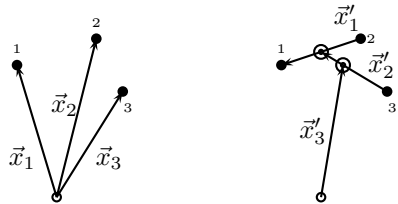
\begin{figure}[!h]
\bc
\psset{unit=1pt}
\begin{pspicture}(150,80)
\pscircle*[linecolor=black](0,50){2} \rput(0,56){\tiny 1}
\pscircle*[linecolor=black](30,60){2} \rput(30,66){\tiny 2}
\pscircle*[linecolor=black](40,40){2} \rput(42,34){\tiny 3}
\pscircle(15,0){2}
\psline{->}(15,0)(0,50)  \rput(2,15){$\vec x_1$}
\psline{->}(15,0)(30,60)  \rput(18,35){$\vec x_2$}
\psline{->}(15,0)(40,40)  \rput(35,15){$\vec x_3$}

\pscircle*[linecolor=black](100,50){2} \rput(100,56){\tiny 1}
\pscircle*[linecolor=black](130,60){2}\rput(134,60){\tiny 2}
\pscircle*[linecolor=black](140,40){2} \rput(142,34){\tiny 3}
\pscircle(115,0){2}
\pscircle[doubleline=true](115,55){3}
\pscircle[doubleline=true](123.33,50){3}
\psline{->}(130,60)(100,50)  \rput(125,68){$\vec x_1'$}
\psline{->}(140,40)(115,55)  \rput(138,50){$\vec x_2'$}
\psline{->}(115,0)(123.33,50)  \rput(112,25){$\vec x_3'$}
\end{pspicture}
\caption{Transformation discussed in the text for 3 particles (black dots). Hollow circles represent the external reference point and double circles represent the particles' total and partial centers of mass.}
\label{ex3}
\ec
\end{figure}

Using, for instance, $a_{11}=1$, $a_{12}=-1$ and $a_{22}=1/2$ one gets for  $\vec x'$ a possible set of Jacobi coordinates
containing the position of the total center of mass, the relative coordinate between particle 1 and 2 and the relative coordinate between particle 3 and the minor center of mass of particles 1 and 2, as shown in Fig. (\ref{ex3}).

\begin{figure}[!h]
\bc
\psset{unit=1pt}
\begin{pspicture}(50,80)
\pscircle*[linecolor=black](00,50){2} \rput(00,56){\tiny 1}
\pscircle*[linecolor=black](30,60){2}\rput(34,60){\tiny 2}
\pscircle*[linecolor=black](40,40){2} \rput(42,34){\tiny 3}
\pscircle(15,0){2}
\pscircle[doubleline=true](23.33,50){3}
\psline{->}(30,60)(00,50)  \rput(25,68){$\vec x_1'$}
\psline{->}(40,40)(30,60)  \rput(48,50){$\vec x_2'$}
\psline{->}(15,0)(23.33,50)  \rput(12,25){$\vec x_3'$}
\end{pspicture}
\caption{A coordinate system that does not obey the set of equations (\ref{sys3}) and thus does not allow for the separation of the center of mass kinetic energy. }
\label{ex3-no}
\ec
\end{figure}
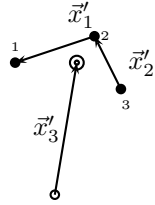

The set of equations (\ref{sys3}) is important, because it tells at a glance that, for example, certain coordinates systems don't have the right property of the separation of the center of mass. Consider the coordinates as in Fig. (\ref{ex3-no}), sometimes called necklace or sequential coordinates \footnote{A folklore note: it is precisely the fact that i was considering the necklace coordinates in connection with many-body nuclear physics models that brought my attention to to the fact that this problem was not thoroughly known and to the formulation of the general conditions under which the separation is possible.}. Using $a_{11}=1$, $a_{12}=-1$ one gets $a_{13}=0$,  $a_{21}=a_{22}$ and therefore $2a_{22}+a_{23}=0$. While the choice of the preceding example works well, the choice $a_{22}=1$ and $a_{23}=-1$ {\it does not satisfy this last equation and therefore does not permit the separation of the center of mass} as described by (\ref{sep}).

The most general matrix for 3 particles is 
\be
A_3(a,b,c)=
\left(
\begin{array}{ccc}
a  & b & -a-b   \\
c & -c\frac{2a+b}{a+2b} & c\frac{a-b}{a+2b} \\
1/3 & 1/3 & 1/3 
\end{array}
\right) 
\label{mat3}
\ee
with arbitrary $a,b,c$. These matrices are not in general a group as before, but remarkably they can still form an additive group under the operation $\pldv$ by restricting some of the parameters. This can be accomplished only in two ways: either $b=ka$ or $a=k'c-2b$, where $k$ and $k'$ are constants. By imposing these two conditions the resulting transformation matrices are respectively 
\be
A_3^{(1)}(a,c)_k=
\left(
\begin{array}{ccc}
a  & ka & -a(1+k)   \\
c & -c\frac{2+k}{1+2k} & c\frac{1-k}{1+2k} \\
1/3 & 1/3 & 1/3 
\end{array}
\right) 
\label{mg1}
\ee
\be
A_3^{(2)}(b,c)_{k'}=
\left(
\begin{array}{ccc}
k'c-2b  & b & b-k'c   \\
c & -2c+\frac{3b}{k'} & c-\frac{3b}{k'} \\
1/3 & 1/3 & 1/3 
\end{array}
\right) 
\label{mg2}
\ee
where the superscripts in parenthesis just denote an arbitrary labeling. The difference between these two matrix groups lies in the fact that one has two arbitrary parameters in the same column, and the other in different columns and, upon reshuffling of the columns, these are the only two possible ways.
It can be noticed that the transformation to Jacobi coordinates of Fig. (\ref{ex3}) is an element of both groups that corresponds to 
\be
A_3(1,-1,1/2)=A_3^{(1)}(1,1/2)_{-1}=A_3^{(2)}(-1,1/2)_{-2} \;.
\ee
An alternative set of Jacobi coordinates is given by $A_3(0,1,-1)$. Notably this is not an element of $A_3^{(1)}$, but it is an element of $A_3^{(2)}(1,-1)_{-2}$. However, by noticing that its matrix corresponds to the previous Jacobi coordinates matrix upon reordering of the columns, one can see that this fact is simply explained by the arbitrariness in choosing the free matrix elements in (\ref{mat3}): we have filled the upper left triangle with arbitrary parameters and determined the rest, but one could as well have made other choices. Summarizing, we must keep in mind that there are two levels of arbitrariness in the procedure of calculation of the matrices: i) it is arbitrary to choose which elements are parameters and which are functions of these parameters and ii) the values of the parameters themselves are arbitrary. 

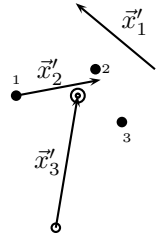
\begin{figure}[!h]
\bc
\psset{unit=1pt}
\begin{pspicture}(50,90)
\pscircle*[linecolor=black](00,50){2} \rput(00,56){\tiny 1}
\pscircle*[linecolor=black](30,60){2}\rput(34,60){\tiny 2}
\pscircle*[linecolor=black](40,40){2} \rput(42,34){\tiny 3}
\pscircle(15,0){2}
\pscircle[doubleline=true](23.33,50){3}
\psline{->}(52.5,60)(22.5,85)  \rput(45,78){$\vec x_1'$}
\psline{->}(0,50)(32,56)  \rput(13,59){$\vec x_2'$}
\psline{->}(15,0)(23.33,50)  \rput(12,25){$\vec x_3'$}
\end{pspicture}
\caption{A rather weird coordinate system corresponding to $A_3(1/2,1,-1)$ that nevertheless allows for the separation of the center of mass kinetic energy.}
\label{ex3-si}
\ec
\end{figure}

Several choices of parameters $a,b,c$ can be traced back to a "rail projection" of a set of Jacobi coordinates, as, for instance, $A_3(1,1,1)$: the first coordinate coincides with $\vec x_1'$ of Fig. (\ref{ex3}),
 the third is the center of mass coordinate $\vec x_3'$ as in all other cases, but the second is just an expansion of  $\vec x_2'$ of Fig. (\ref{ex3}) along the two rails connecting the reference point with $3$ and with the center of mass of particles $1$ and $2$. This is analogous to the case already discussed for 2 particles.
Apart from these cases, there are others that, despite their strange look, are equally good at separating out exactly the kinetic energy of the center of mass. Take for example $A_3(1/2,1,-1)$ that gives the coordinate transformation depicted in Fig. (\ref{ex3-si}).
Albeit this might be useless to the solution of the many-body problem, it is a viable alternative to Jacobi coordinates. This example shows clearly that, when $n>2$, the group structure has a geometrical interpretation that is wider than the parallel expansion/contraction along the rails described above, but several subsets of all possible transformations still have this interpretation. In particular for the matrix (\ref{mg1}) it is possible to find fixed rails, because $a$ and $c$ amount to a multiplicative constant for $\vec x_1'$ and $\vec x_2'$ respectively, but in this case, by changing $k$, one is not only shifting the vector's tips along the rails, but it's also changing its direction accordingly. The interpretation of (\ref{mg2}) is instead less obvious, because two parameters enter at the same time into several matrix elements.

For $n$ particles we return to the system (\ref{co1}). The corresponding transformation matrix that maps the initial set of coordinates into a set of coordinates that allows the exact separation of the center of mass kinetic energy can be schematically divided in the following way:
\be
A_n(a_1, \cdots, a_p) =
\label{matri}
\ee
\bc
\psset{unit=1.08pt}
\begin{pspicture}(80,80)
\rput(40,40){ $
\left(
\begin{array}{cccc}
a & \cdots  &  & c_2 \\
\vdots &  &  & \\
& & \vdots &  \vdots \\
 & \cdots & c_1 &  \\
~1/n & \cdots &  & 
\end{array}
\right) $}
\pspolygon[linestyle=dotted, dotsep=1](60,76)(11,76)(11,17)(19,17)(60,68)
\pspolygon[linestyle=dotted, dotsep=1](60,64)(60,17)(22,17)
\pspolygon[linestyle=dotted, dotsep=1](63,76)(76,76)(76,17)(63,17)
\pspolygon[linestyle=dotted, dotsep=1](11,15)(76,15)(76,5)(11,5)
\end{pspicture}
\ec
where the $n$ matrix elements of the last row are all equal to $1/n$. Conditions (\ref{co1}-2) is used to determine, for example, the remaining $n-1$ elements of the last column (one could have equivalently chosen another column):
\be
a_{in}=-\sum_{k=1}^{n-1} a_{ik} \qquad i=\{1,\cdots, n-1 \} \;.
\ee 
Conditions (\ref{co1}-1) are used to set the $(n-1)(n-2)/2$ matrix elements of the lower triangle as functions of the arbitrary $p=n(n-1)/2$ matrix elements of the upper triangle, starting from the pair $\{i,j\}=\{1,2\}$
and proceeding row by row.
Matrices (\ref{matri}) most certainly don't form a group under $\pldv$, but it is still possible that, as in the case $n=3$ they would do so by restricting the number of parameters under certain special conditions.

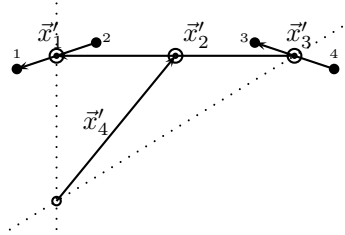
\begin{figure}[!h]
\bc
\psset{unit=1pt}
\begin{pspicture}(120,90)(-10,-10)
\pscircle*[linecolor=black](00,50){2} \rput(00,56){\tiny 1}
\pscircle*[linecolor=black](30,60){2}\rput(34,62){\tiny 2}
\pscircle*[linecolor=black](90,60){2} \rput(86,62){\tiny 3}
\pscircle*[linecolor=black](120,50){2} \rput(120,56){\tiny 4}
\pscircle(15,0){2}
\pscircle[doubleline=true](60,55){3}
\pscircle[doubleline=true](105,55){3}
\pscircle[doubleline=true](15,55){3}
\psline{->}(30,60)(0,50)  \rput(107,63){$\vec x_3'$}
\psline{->}(120,50)(90,60)  \rput(13,63){$\vec x_1'$}
\psline{->}(105,55)(15,55)  \rput(68,63){$\vec x_2'$}
\psline{->}(15,0)(60,55)  \rput(30,30){$\vec x_4'$}
\psline[linestyle=dotted,dotsep=3pt](15,-11)(15,75)
\psline[linestyle=dotted,dotsep=3pt](-3,-11)(123,66)
\end{pspicture}
\caption{Coordinate system corresponding to $A_4(1,-1,0,1/2,1/2,0)$. The rails are indicated as dotted lines. This system can be used to model scattering of two deuterons for example.}
\label{ex4}
\ec
\end{figure}

Let us give an example of the usefulness of the conditions (\ref{co1}). With reference to Fig. (\ref{ex4}), suppose that we wish to use the internal coordinates of the two subsystems formed by particle $1-2$ and $3-4$ and in addition we want also the right property of separating the total kinetic energy center of mass (into the new coordinate $\vec x_4'$).

By filling up the transformation matrix for four particles with the proper values for the first, third and fourth row, we get
\be
\left(
\begin{array}{cccc}
1  & -1 & 0 &0    \\
a & a & -a & -a \\
0 &0 & 1 & -1 \\
1/4 & 1/4 & 1/4 & 1/4 
\end{array}
\right) 
\ee
therefore the entire class of possible transformations in this particular case reduces to all the vectors 
with direction parallel to that the center of mass of $3-4$ to the center of mass of $1-2$, or, in other words all expansions/contractions along the rails shown in figure as dotted lines. With the choice $a=1/2$ we have the simplest of these vectors, i.e. $\vec x_2'$ of Fig. (\ref{ex4}). 
This coordinates are suitable to describe, for instance, the scattering of two deuterons, where the total wave function can be expanded in terms of products like $\phi_{12}(\vec x_1')\phi_{34}(\vec x_3')\psi(\vec x_2')$,
where $\phi$ are internal wavefunctions that describe separately the subsystems $1-2$ and $3-4$ and $\psi$ is the relative motion wavefunction between the two subsystem. Several other systems might be treated with  coordinates of this type: cluster states in light nuclei, tetratomic molecules and so on.

In conclusion, we have discussed the most general conditions to impose on coordinates transformations that allow for the exact separation of the total center of mass kinetic energy, obtaining the set of equations (\ref{cond}).
These equations show that there are classes of coordinates systems that can be used in place of usual Jacobi coordinates, achieving the same goal of elimination of the center of mass energy.
The application to small numbers of particles has been discussed and we have found that a curious additive group structure is hidden in these transformation matrices. We have analyzed it finding a connection with a simple geometrical interpretation. We believe that these results can be of some importance to quantum many-body models based on the solution of the non-relativistic Schr\"odinger equation and that the new group structure might reveal further hidden symmetries and analogies that might help advancing this field.

\end{document}